\documentclass[aps,prd,reprint,superscriptaddress,nofootinbib,amsfonts,amssymb,amsmath]{revtex4-1}

\usepackage[final]{graphicx}
\usepackage{hyperref}
\usepackage{amsmath}
\usepackage{bbm}
\usepackage{amsfonts}
\usepackage{amssymb}
\usepackage{latexsym}
\usepackage{graphicx}
\usepackage[english]{babel}
\usepackage{multirow}
\usepackage{float}
\usepackage{url}
\usepackage{slashed}
\usepackage{xcolor}
\usepackage[utf8]{inputenc}
\usepackage{lipsum}

\usepackage[utf8]{inputenc}

\usepackage{diagbox}

\usepackage{mathtools}
\usepackage{amsfonts}
\usepackage{mathrsfs}
\usepackage{bbm}
\usepackage{slashed}

\usepackage{graphicx}
\usepackage{color}
\usepackage{array}
\usepackage{esint}
\usepackage{placeins}
\usepackage{booktabs}
\usepackage{makecell}
\usepackage{epstopdf}
\usepackage[caption=false]{subfig}

\usepackage{xspace}
\usepackage{siunitx}
\usepackage{hyperref}
\usepackage[nameinlink]{cleveref}
\usepackage{appendix}

\usepackage{xifthen}
\usepackage{xcolor}
\hypersetup{
	colorlinks,
	linkcolor={red!75!black},
	citecolor={blue!75!black},
	urlcolor={blue!75!black}
}
\usepackage{comment}

\setkeys{Gin}{width=0.48\textwidth}
\graphicspath{
	{./figures/}
	{../figures/}
}

\def\s0#1#2{\mbox{\small{$ \frac{#1}{#2} $}}}
\def\0#1#2{\frac{#1}{#2}}

\usepackage{xcolor}

\usepackage{ulem}

\usepackage{amsmath,mathrsfs,braket,hyperref,bm,amssymb}
\usepackage {slashed}
\usepackage{graphicx} 

\usepackage{tikz}
\usepackage{tikz-feynman}

\newcommand{\bmu}{\bar\mu}
\newcommand{\gammaE}{{\gamma_\rmii{E}}}

\newcommand{\be}{\begin{equation}}
\newcommand{\ee}{\end{equation}}
\newcommand{\lp}{\left(}
\newcommand{\rp}{\right)}
\newcommand\diff{\,\mathrm{d}}
\newcommand{\MSbar}{\overline{\rm{MS}}}

\newcommand{\nn}{\nonumber}

\newcommand{\rmii}[1]{{\mbox{\tiny\rm{#1}}}}

\begin{document}
\newcommand{\CQU}{\affiliation{Department of Physics and Chongqing Key Laboratory for Strongly Coupled Physics, Chongqing University,Chongqing 401331, P. R. China}}

\title{Gauge-Invariant Bubble Nucleation and Gravitational Waves from First-Order Electroweak Phase Transitions}

\author{Jie Liu}
\affiliation{~Department of Physics and Chongqing Key Laboratory for Strongly Coupled Physics, Chongqing University, Chongqing 401331, P. R. China}

\author{Renhui Qin}
\email{20222701021@stu.cqu.edu.cn}
\affiliation{~Department of Physics and Chongqing Key Laboratory for Strongly Coupled Physics, Chongqing University, Chongqing 401331, P. R. China}

\author{Ligong Bian}
\email{lgbycl@cqu.edu.cn}
\affiliation{~Department of Physics and Chongqing Key Laboratory for Strongly Coupled Physics, Chongqing University, Chongqing 401331, P. R. China}

\begin{abstract}
A long-standing problem in electroweak phase transition studies is the spurious gauge dependence of bubble nucleation rate calculations, which introduces large systematic uncertainties in predictions of gravitational waves and related cosmological phenomena. We address this issue by presenting a systematic study of gauge-invariant bubble nucleation within a phenomenologically realistic effective field theory framework. Using the three-dimensional thermal effective field theory formalism with a consistent power-counting scheme, we rigorously establish the gauge invariance of the physical nucleation rate up to two-loop order. This proof eliminates the gauge-parameter dependence of key phase transition parameters, providing a robust theoretical foundation for precise calculations of gravitational waves, electroweak baryogenesis, and primordial magnetogenesis.
\end{abstract}

\maketitle

\noindent{\it Introduction.}
The persistent challenge of an unphysical gauge dependence in the bubble nucleation rate has for decades seriously undermined the reliability and predictive power of perturbative theoretical calculations within thermal effective field theory and complicated phenomenological applications derived from thermal phase transition studies,  such as the prediction of the primordial magnetic field
(PMF) to address the source of the cosmological magnetic field~\cite{Yang:2021uid,Di:2020kbw,Zhang:2019vsb,Stevens:2012zz,Kahniashvili:2009qi,Hindmarsh:1997tj,Grasso:1997nx,Ahonen:1997wh}, electroweak baryogenesis to intereprate the baryon asymmetry of the universe~\cite{Morrissey:2012db,Cohen:1993nk}, and gravitational wave (GW) predictions\cite{Croon:2020cgk,Athron:2023xlk,Qin:2024dfp,Zhu:2025pht} that are preliminary scientific target of Laser Interferometer Space Antenna (LISA)~\cite{LISA:2017pwj,Baker:2019nia},  Taiji~\cite{Hu:2017mde,Ruan:2018tsw}, and TianQin~\cite{TianQin:2015yph,Zhou:2023rop}.

The bubble nucleation rate describes a fundamental physical process and should not depend on an artificial gauge choice~\cite{Patel:2011th}. Fortunately, the Nielsen identity provides the chance to obtain gauge-independent quantities from the gauge-dependent action and effective potential~\cite{Hu:1996qa,Nielsen:1975fs}. The Nielsen identities are argued to guarantee the gauge-independence order-by-order, and previous studies have justified this statement in the abelian Higgs model at finite temperature~\cite{Garny:2012cg,Metaxas:1995ab,Arunasalam:2021zrs,Hirvonen:2021zej,Lofgren:2021ogg}. Ref.~\cite{Hirvonen:2021zej,Lofgren:2021ogg} provide the first proof of the gauge invariant of the nucleation in the Abelian Higgs model.Refs.~\cite{DiLuzio:2014bua,Espinosa:2016nld} determined the gauge-independent scale of the Standard Model vacuum instability. However, to the best of our knowledge, no study within realistic gauge-Higgs theories has yet been conducted to address the uncertainties in predicting vacuum decay via a first-order electroweak phase transition. Our work aims to bridge this gap.

In this Letter, for the first time, we generalize the Nielsen-identity-based proof of bubble nucleation rate gauge invariance from the Abelian Higgs model to the Non-Abelian $SU(2)_L\times U(1)_Y$ electroweak theory. This work constitutes a critical prerequisite for reliable quantitative predictions of phase transition dynamics and associated gravitational wave signals, removing a long-standing systematic uncertainty that has plagued the field for decades. The electroweak phase transition in the Standard Model is found to be a crossover through lattice simulation~\cite{DOnofrio:2014rug}. A first-order phase transition necessitates new physics (NP) beyond the Standard Model~\cite{Athron:2023xlk, Caldwell:2022qsj}. The Standard Model effective field theory (SMEFT), which provides a powerful bottom-up effective theory approach to probe the new physics beyond the Standard Model~\cite{Ellis:2018gqa,Ellis:2020unq,Ethier:2021bye}, also validates a strong first-order PT detectable by near-future GW observatories~\cite{Ellis:2018mja,Cai:2017tmh,Croon:2020cgk,Grojean:2004xa,Delaunay:2007wb,Chala:2018ari,Bodeker:2004ws,Damgaard:2015con,Hashino:2022ghd,Postma:2020toi,Chala:2025aiz,Camargo-Molina:2024sde}. Additionally, by applying dimensional reduction (DR) to the four-dimensional (4D) theory to construct a three-dimensional (3D) effective field theory, the Linde problem of perturbative calculations can be avoided~\cite{Linde:1980ts,Gross:1980br}. This approach enables lattice simulations to determine the phase transition properties~\cite{Farakos:1994xh,Kajantie:1995kf,Niemi:2024axp,Niemi:2020hto,Andersen:2017ika,Annala:2025aci} and the bubble nucleation rate~\cite{Moore:2000jw,Moore:2001vf,Gould:2024chm,Gould:2022ran}. Therefore, all calculations in this Letter are performed within the 3D SMEFT, obtained via DR procedure from the 4D SMEFT with a consistent power-counting scheme.

\noindent{\it The 3D Model.}
From the DR strategy based on the chain of scale hierarchies at finite temperature, one can systematically integrate out fermions{\color{blue}{,}} and temporal scalars of gauge bosons{\color{blue}{,}} and arrive at the high temperature effective field theory (3D EFT) to study the thermodynamics~\cite{Kajantie:1995dw,Braaten:1995cm}, such as Higgs condensation and phase transition parameters. Within the framework under study, at the ultra-soft scale, we have
\begin{equation}
    \begin{aligned}
\mathcal{L}^{\mathrm{ultrasoft}}_{3d}=&-\frac{1}{4}W^a_{ij}W^a_{ij}-\frac{1}{4}B_{ij}B_{ij}+(D_i H)^\dagger(D_i H)\\
&+\mathcal{L}_\mathrm{g.f}+\mathcal{L}_{\mathrm{ghost}}
-V^{\mathrm{ultrasoft}}_{3d}\;.
    \end{aligned}
\end{equation}
Here $i,j$ are spatial indices $(i,j=1,2,3)$. The implicit gauge couplings are $\bar{g}_3$ for the SU(2) and $\bar{g}^{\prime}_3$ for the U(1) sectors. The ultra-soft scale potential reads
\begin{equation}
   V^{\mathrm{ultrasoft}}_{3d}=-\bar{m}_3^2 H^\dagger H+\bar{\lambda}_3 (H^\dagger H)^2+\bar{c}_{6,3} (H^\dagger H)^3\;.
\end{equation}
Here, the last term makes it possible to realize a first-order electroweak phase transition by including new physics (NP) effect with a dimension-six operator. The relation between $\bar{c}_{6,3}$ in the 3D EFT and $c_6 \equiv 1/\Lambda^2$ in the 4D SMEFT can be found in Ref.~\cite{Croon:2020cgk}, with $\Lambda$ being the NP scale. This operator captures the main effect of the SMEFT to accommodate the first-order electroweak phase transition.
In this study, we adopt the background field $R_\xi$ gauge, since the IR divergences would afflict Fermi gauge ~\cite{Biekotter:2025npc,Garny:2012cg} in the context of the Nielsen identity~\cite{Nielsen:1975fs,Aitchison:1983ns}.
The gauge-fixing term is:
\begin{eqnarray}\label{eq:soft g.f}
\mathcal{L}^{\mathrm{ultrasoft}}_{\mathrm{g.f}}&=&-\frac{1}{2 \xi}\lp\partial_i A^{a}_i+i\frac{\bar{g}_3}{2} \xi \lp H^\dagger \sigma^a \Phi_0-\Phi_0^\dagger \sigma^a H\rp \rp^2\,\nonumber\\
		&-&\frac{1}{2 \xi}\lp \partial_i B_i+i\frac{\bar{g_3^\prime}}{2} \xi\lp H^\dagger \Phi_0-\Phi_0^\dagger H \rp\rp^2\;,
\end{eqnarray}
and the ghost part reads:
\begin{equation}
	\mathcal{L}^{\mathrm{ultrasoft}}_{ghost}=-\begin{pmatrix}
		\bar{c}^a & \bar{c}^0
	\end{pmatrix} \begin{pmatrix}
		M^{ab} & M^a\\
		M^b & M
	\end{pmatrix}\begin{pmatrix}
		c^b\\
		c^0
	\end{pmatrix}\;,
\end{equation}
with
\begin{equation}
	\begin{aligned}
		M^{ab}&=(\partial^i D_i^{ab})+\bar{g}_3^2\xi[(t^b H)^\dagger(t^a \Phi_0)+(t^a \Phi_0)^\dagger(t^b H)]\,,\\
		M^a&=\frac{\bar{g}_3 \bar{g_3^\prime}}{2}\xi [H^\dagger t^a \Phi_0+(t^a \Phi_0)^\dagger H]\,,	\\
		M^b&=\frac{\bar{g}_3 \bar{g_3^\prime}}{2}\xi [(t^b H)^\dagger \Phi_0+\Phi_0^\dagger (t^b H)]=M^a\,,\\
		M&=\partial^2+\frac{\bar{g_3^\prime}^2}{4}(H^\dagger \Phi_0+\Phi_0^\dagger H)\;,
	\end{aligned}
\end{equation}
where $ D_i^{ab}=\partial_i- \bar{g}_3 f^{abc} A^c_i$.
In the following, we work at the ultra-soft scale with the power-counting of the weak gauge coupling $\lambda\sim g^3$ and drop the superscripts and subscripts of bosonic fields, weak gauge couplings, quartic couplings, and those of $c_6$ for convenience, unless explicitly stated.

With the Higgs doublet in the 3D Lagrangian density expressed as
$H=(
		\chi^1(x) + i \chi^2(x) ,\phi + h(x) + i \chi^3(x)
)^T/\sqrt{2}
$,
and considering the gradient expansion in powers of spatial derivatives, the effective Euclidean action reads~\cite{Linde:1981zj,Moss:1985ve}:
\begin{equation}\label{eq:action3d}
   S_3= \int \diff^3 x\Bigl[ V^\mathrm{eff}(\phi,T) +\frac{1}{2}Z(\phi,T) \lp\partial_i \phi\rp^2 +\cdots \Bigr]\;,
\end{equation}
where $Z$ is the renormalization factor of the wave function:
\begin{align}
\label{eq:Zexp}
  Z &= 1
    + Z_{\mathrm{NLO}}
    + \mathcal{O}(g^2)
    \;,
\end{align}
where, we have
\begin{equation}
     Z_{\mathrm{NLO}}= -\frac{11 \left(\sqrt{g^2+g^{'2}}+2 g\right)}{48 \pi  \phi }\;.\label{znlo}
\end{equation}

At the ultra-soft scale, we have
\begin{equation}\label{Veff}
V^\mathrm{eff}(\phi,T)=V_{\rm LO}+V_{\rm NLO}\;,
\end{equation}
with the leading order contribution being
\begin{equation}\label{VLO}
    V_{\mathrm{LO}}=-\frac{1}{2}m^2\phi^2+\frac{\lambda \phi^4}{4}+\frac{c_6 \phi^6}{8}-\frac{1}{12 \pi}\lp 4m_W^3+2m_Z^3\rp\,,
\end{equation}
when the relation between the scalar coupling and weak gauge coupling $\lambda\sim g^3$ holds~\cite{Arnold:1992rz,Liu:2026ask}.
The dimension-six operator here is understood as modifying the mass and coupling parameters at tree-level~\cite{Andersen:2017ika,Gorda:2018hvi,Laine:1996ms} rather than from the DR procedure, whose effect is negligible~\cite{Farakos:1994kx}.
The last term comes from one-loop corrections of gauge bosons with the field-dependent masses given in Supplemental Material. Since $m_G^2=-m^2+\lambda \phi^2+\frac{3}{4}c_6\phi^4 \sim 0$, we have
$(m_G^2+m_c^2)^\frac{3}{2}-m_c^3 \sim \frac{3}{2} m_G^2 m_c\,,$
therefore, the Goldstone and ghost terms' contribution at one-loop are counted as next-to-leading order (NLO).
We find the effective thermal potential at NLO can be separated into two parts:
\begin{equation}
  V_{\mathrm{NLO}}=V^0_{\mathrm{NLO}}+\sqrt{\xi }V^\xi_{\mathrm{NLO}}\;, \label{vnlo}
\end{equation}
where $V^0_{\mathrm{NLO}}$ does not depend on the gauge choice parameter $\xi$, and all the $\xi-$dependent terms are collected in $V^\xi_{\mathrm{NLO}}$, which is calculated as:
\begin{equation}
    V^\xi_{\mathrm{NLO}}=\frac{(2m_W+m_Z)(2 m_W^3+m_Z^3)}{16 \pi^2 \phi^2}-\frac{2 m_W+m_Z}{8 \pi}m_G^2\;,
\end{equation}
the last term comes from the one-loop Goldstone-ghost contribution; the rest comes from two-loop contributions.
The $V^0_{\mathrm{NLO}}$ takes the form:
\begin{equation}
\begin{aligned}
V^0_{\mathrm{NLO}}&=A \phi^2+B \phi^2 \ln (\frac{\phi}{\bmu})\,,
\end{aligned}
\end{equation}
with $\bmu$ as the renormalization scale, the uncertainty from renormalization scale dependence is negligible at two-loop order in the dimensional reduction approach~\cite{Croon:2020cgk,Qin:2024idc}. At the order of $\sim \mathcal{O}(g^4)$, the first term in $V^0_{\mathrm{NLO}}$ reads,
\begin{eqnarray}
A&=&\frac{c_W^3 \left(36 g^6+28 g^4 g^{\prime 2}-15 g^2 g^{\prime 4}-4 g^{\prime 6}\right)}{1024 \pi ^2 g^2}\nonumber\\
&+&\frac{c_W^2 \left(71 g^6+20 g^4 g^{\prime 2}-78 g^2 g^{\prime 4}-15 g^{\prime 6}\right)}{3072 \pi ^2 g^2}\nonumber\\
&+&\frac{c_W^4 \left(5 g^4+9 g^2 g^{\prime 2}+g^{\prime 4}\right) \log (g)}{1024 \pi ^2}\nonumber\\
&+&\frac{c_W^4 \left(20 g^4 g^{\prime 4}+19 g^2 g^{\prime 6}+2 g^{\prime 8}\right) \log \left(\frac{g}{c_W}+g\right)}{1024 \pi ^2 g^4}\nonumber\\
&+&\frac{\left(3 g^4+14 g^2 g^{\prime 2}+3 g^{\prime 4}\right) \log \left(\frac{g}{c_W}\right)}{1024 \pi ^2}\nonumber\\
&+&\frac{c_W^2 \left(-42 g^6-19 g^4 g^{\prime 2}+8 g^2 g^{\prime 4}+g^{\prime 6}\right) \log \left(\frac{g}{c_W}+2 g\right)}{1024 \pi ^2 g^2}\nonumber\\
&+&\frac{c_W^2 \log (2) \left(13 g^6+9 g^4 g^{\prime 2}\right)}{256 \pi ^2 g^2}\;,
\end{eqnarray}
with the weak mixing angle $\cos\theta_W\equiv c_W=g/\sqrt{g^2+g^{\prime 2}}$, and
\begin{equation}
    B=\frac{-17 g^4+18 g^2 g^{\prime2}+3 g^{\prime4}}{512 \pi ^2}\;.
\end{equation}
The detailed derivation of all the two-loop contributions are provided in the companion article~\cite{Liu:2026ask}, the additional topological Feynman diagram contributions in comparison with the Landau gauge~\cite{Croon:2020cgk} are given in the Supplemental Material.

For the gauge dependence analysis in the high temperature dimensional reduction, one needs to require $|\xi| \ll 1/g$ to preserve the perturbativity of the 3D EFT~\cite{Croon:2020cgk,Kripfganz:1995jx,Laine:1994bf,Garny:2012cg}, which means $|\xi|\lesssim \mathcal{O}(1)$ and justifies our $\xi-$expansion analysis. We further note that this situation can yield results close to those of the Landau gauge~\cite{Wainwright:2011qy,Karjalainen:1996rk,Laine:1994zq}.

 The precise determination of the first-order phase transition dynamic crucially depend on the bubble nucleation rate~\cite{Coleman:1977py,Linde:1981zj}. The inclusion of the typical scale of the nucleated bubbles allows a perturbative study of vacuum decay through bubble nucleation~\cite{Gould:2021ccf}. The bubble nucleation rate per Hubble volume, $\Gamma$, admits the semiclassical approximation~\cite{Coleman:1977py,Linde:1981zj}
\begin{equation}
\Gamma = A(T) e^{-\mathcal{B}}\;.
\end{equation}
Here, the exponent $\mathcal{B}$ encodes the leading behavior of the rate and can be evaluated from the ``bounce'' solution after solving the classical Euclidean field equations, with $\mathcal{B}\equiv S_3|_{\phi=\phi_b}$.
The prefactor $A(T)$ includes the statistical ($A_{\rm stat}$) and dynamical ($A_{\rm dyn}$) factors for thermal bubble nucleation, $A_{\rm stat}$ can be calculated based on Gelfand-Yaglom theorem~\cite{Wang:2026ifp,Athron:2023xlk,Ekstedt:2023sqc}, and
$A_{\rm dyn}$ can be extracted from real time simulations of the nucleation rete~\cite{Hirvonen:2024rfg,Hirvonen:2025hqn,Wang:2025ooq}.
At the nucleation temperature $T_n$, the bubble nucleation rate equals the Hubble parameter, $\Gamma\sim H$, corresponding to $\mathcal{B}\approx 140$.

Traditionally, one can straightforwardly solve the bounce function obtained after applying the principle of least action from Eq.~\ref{eq:action3d}.
Since the $d Z/d\phi$ factor at the vacuum expectation value (VEV) is close to zero, the bounce function becomes~\cite{Qin:2024dfp}:
\begin{equation}\label{bouncefunction}
\frac{d^2 \phi_b}{d\rho^2}+\frac{2}{\rho}\frac{d\phi_b}{d\rho}=\frac{1}{Z}\frac{d V^{\mathrm{eff}}}{d\phi}\;,
\end{equation}
and we use the code ``FindBounce'' to solve this equation and obtain the nucleation temperature $T_n$ and the field configuration of the bounce solution~\cite{Guada:2020xnz},
with $\phi_b(\infty)=C_{\infty}K_{1/2}(|m| \rho)|\frac{m}{\rho}|^{1/2}$, where $K_{1/2}(x)$ is the second kind modified Bessel function and $C_{\infty}$ is a constant~\cite{Wang:2026ifp}.

The PT temperature and the duration determine the peak frequency of the produced GW from PT \cite{Huber:2008hg,Caprini:2015zlo,Caprini:2009yp}, and the trace anomaly $\alpha$ usually determines the amplitude of the generated GW. The inverse duration of the PT is defined as $\beta/H_n=T_n(dS_3/dT)|_{T_n}$, and the  $\alpha$ is defined as $\alpha=T_n\Delta\rho/\rho_{rad}$ with
\begin{equation}\label{deltarho}
\Delta \rho=\left.-\frac{3}{4}\Delta V(\phi_n,T_n)+\frac{1}{4}T_n\frac{d\Delta V(\phi_n,T)}{dT}\right|_{T=T_n},
\end{equation}
where $\Delta V(\phi_n,T)=V^{\mathrm{eff}}(\phi_n,T)-V^{\mathrm{eff}}(0,T)$ denotes the potential difference between the false and true vacuum of $\phi_n$ at $T_n$, and $\rho_{rad}=\pi^2g_* T^4_n/30$, $g_*=106.75$ is the effective number of relativistic degrees of freedom \cite{Croon:2020cgk}.

\noindent{\it Gauge-invariant nucleation rate.}
Consider the power counting of the weak gauge coupling $g$, and from Eqs.~(\ref{eq:action3d},\ref{eq:Zexp},\ref{Veff}), we can expand the exponent of the nucleation rate as $\mathcal{B} = \mathcal{B}_0 + \mathcal{B}_1$:
\begin{align}
\label{eq:B0}
\mathcal{B}_0 &=
   \int \diff^3 x \left[
      V_{\rmii{LO}}(\phi_b)
    + \frac{1}{2}\left(\partial_i \phi_b\right)^2
  \right]
  \;,\\
\label{eq:B1}
\mathcal{B}_1 &=
   \int \diff^3 x \left[
      V_{\rmii{NLO}}(\phi_b)
    + \frac{1}{2}Z_{\rmii{NLO}}(\phi_b)\left(\partial_i \phi_b\right)^2
  \right]
  \;,
\end{align}
up to the $g^3$ and $g^4$ orders respectively.
Where, $\phi_b(x)$ is the bounce solution of Eq.~\ref{eq:B0}.
According to the power-counting analysis method developed in Ref.~\cite{Gould:2021ccf}, the characteristic scale of bubble nucleation is
related to the effective mass of the nucleating field, i.e., $\sim g^{\frac{3}{2}}$. In this scheme, the bubble nucleation rate reads
\begin{align}
\label{eq:rate-LO}
    \Gamma=A(T)e^{-\mathcal{B}_0-\mathcal{B}_1}=A' e^{-\lp a_0 g^{-\frac{3}{2}} + a_1 g^{-\frac{1}{2}}\rp}\;,
\end{align}
with $a_{0,1}$ being numerical coefficients that can be computed directly from the gradient expansion at LO and NLO, respectively.
It should be noted that the derivative expansion is valid, in the sense that the expansion in the power-counting scheme $\lambda\sim g^3$ satisfies $P^2/M^2\sim g$, and therefore, $\mathcal{B}_0\sim g^{-3/2}$, and $\mathcal{B}_1\sim g^{-1/2}$~\cite{Hirvonen:2021zej}.
The prefactor $A'$ encodes the prefactor of $A(T)$ and other higher-order corrections, which we leave for future study. Our goal is to compute the LO and NLO terms in $\mathcal{B}$, i.e., $\mathcal{B}_{0,1}$, and show their gauge invariance. This then ensures the gauge invariance of the nucleation rate $\Gamma$ and, further, of the PT parameters provided the prefactor $A'$ is gauge independent. The prefactor was shown to be gauge independent in the SU(2) Higgs model at finite temperature~\cite{Baacke:1999sc,Kierkla:2025qyz} and in the 4D Abelian Higgs model~\cite{Endo:2017gal}. Therefore, we treat the $A^\prime$ as an overall constant in analogy with that of Ref.~\cite{Lofgren:2021ogg}.

We now apply the procedure of the Nielsen identity to the 3D thermal effective field theory after performing the DR porcedure, and justify the gauge invariance of $\mathcal{B}$ through the Nielsen identity. Since $\mathcal{B}_{0}$ is gauge independent, as follows from Eqs.~\ref{eq:B0} and ~\ref{VLO}, we focus on proving the gauge invariance of $\mathcal{B}_{1}$.
Firstly, the gauge-dependent effective action satisfies the following equation~\cite{Metaxas:1995ab}:
\begin{equation}
    \xi\frac{\partial }{\partial \xi}S[\phi(x),\xi]=-\int_y K[\phi(y),\xi]\frac{\delta S}{\delta \phi(y)}\;,
\end{equation}
where the $K[\phi]$ admits a derivative expansion:
\begin{equation}
    K[\phi]=C(\phi)+D(\phi)(\partial_\mu \phi)^2-\partial_\mu(\tilde{D}(\phi)\partial_\mu \phi)+\mathcal{O}(\partial^4)\,.
\end{equation}
Consider the expansion of the effective action, this yields the Nielsen identities for the effective potential and the field renormalization factor
\begin{eqnarray}
  \label{Nielsen_1}
&&\xi\frac{\partial V}{\partial \xi}=-C \frac{\partial V}{\partial \phi}  \;,\\
&&\xi\frac{\partial Z}{\partial \xi}=-C\frac{\partial Z}{\partial \phi}-2 Z \frac{\partial C}{\partial \phi}-2 D \frac{\partial V}{\partial \phi}-2 \tilde{D}\frac{\partial^2 V}{\partial \phi^2}\;.\label{Nielsen_2}
\end{eqnarray}
To employ these relations, we expand them in powers of the weak gauge coupling $g$ by first noting the scaling of the Nielsen coefficients
$C=C_g+C_{g^2}+\cdots\ $, and $D, \tilde{D}=\mathcal{O}(g^{-1})$, see the Supplemental Material and the companion article~\cite{Liu:2026ask} for details. The function of $C$ at the order of $\mathcal{O}(g)$ reads
\begin{equation}
    C_{\mathrm{LO}}=\frac{\lp 2 g+\sqrt{g^2+g^{'2}}\rp \sqrt{\xi}}{32 \pi}\,.\label{CLO}
\end{equation}
With Eqs.~(\ref{CLO},\ref{znlo},\ref{VLO},\ref{vnlo}) at hand, we find that,  up to order $\sqrt{\xi}$, the leading order of Nielsen identities, i.e., at the powers of $\mathcal{O}(g^4)$ in the first Nielsen identity~\eqref{Nielsen_1} and $\mathcal{O}(g)$ in the second Nielsen identity \eqref{Nielsen_2}, yields
\begin{align}
\label{eq:highT-nielsen-1}
 \xi \frac{\partial}{\partial \xi} V_{\rmii{NLO}} &=
  - C_{\rmii{LO}} \frac{\partial}{\partial \phi} V_{\rmii{LO}}
  \;,\\
\label{eq:highT-nielsen-2}
\xi \frac{\partial}{\partial \xi} Z_{\rmii{NLO}} &=
  -2  \frac{\partial}{\partial \phi} C_{\rmii{LO}}
  \;.
\end{align}
With above Nielsen identities, we can demonstrate the gauge independence of $\mathcal{B}_1$ provided that the bounce solution is valid at leading order when one count $\lambda\sim g^3$:
\begin{eqnarray}
\xi \frac{\partial }{\partial \xi}\mathcal{B}_1&=&\xi \frac{\partial }{\partial \xi}\int \diff^3 x \left[ V_{\rm NLO}(\phi_b)+\frac{1}{2}Z_{\rm NLO}(\phi_b)(\partial_\mu \phi_b)^2 \right]\,\nonumber\\
&=&\int \diff^3 x \left[-C_{\rm LO}\frac{\partial V_{\rm LO}}{\partial \phi}-\frac{\partial C_{\rm LO}}{\partial \phi}(\partial_\mu \phi_b)^2\right]\,\nonumber\\
&=&-\int \diff^3 x \left[C_{\rm LO}\frac{\partial V_{\rm LO}}{\partial \phi}+\partial^\mu C_{\rm LO} (\partial_\mu \phi_b)\right]\,\nonumber\\
&=&-\int \diff^3 x C_{\rm LO} \left[\frac{\partial V_{\rm LO}}{\partial \phi}-\Box \phi_b\right]=0\;.\label{B1xi}
\end{eqnarray}
This formula can be recognized as generalizations of
the corresponding equations from the Abelian-Higgs model~\cite{Lofgren:2021ogg,Hirvonen:2021zej} to electroweak theories.

\begin{figure}[!htp]
    \centering
    \includegraphics[width=0.7\linewidth]{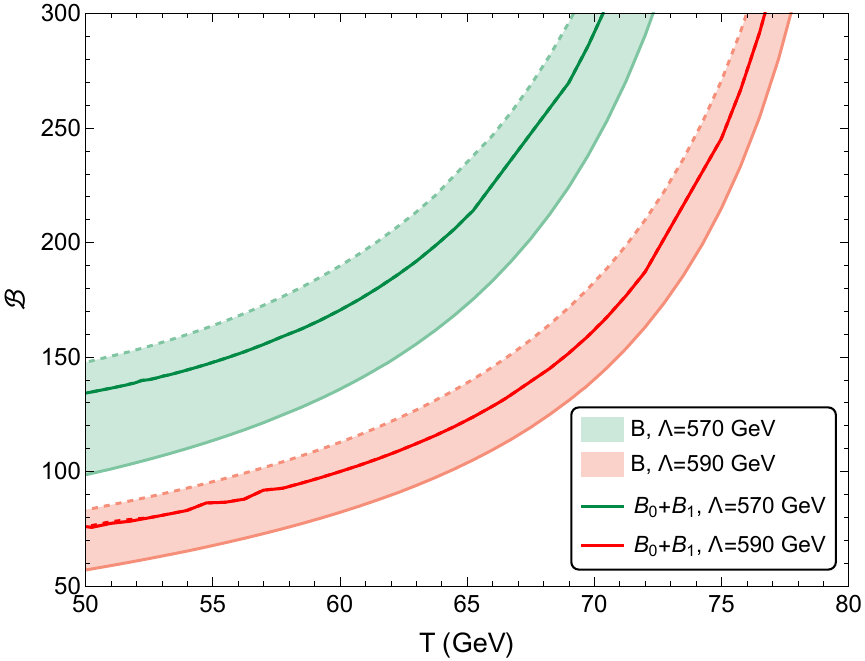}
    \caption{The effective action $\mathcal{B}$ as a function of temperature $T$ for gauge parameter $\xi\in[0,1]$ with NP scales $\Lambda=570$ GeV and $590$ GeV. Lighter-colored curves denotes the results of tradition method with the solid and dashed lines representing $\xi = 0$ and $\xi = 1$, respectively, while darker solid curves denotes the results from the gauge-invariant framework.
    }
    \label{fig:S3}
\end{figure}

We now investigate the gauge dependence of the bounce action. We perform numerical calculations with the gauge parameter
$\xi\in[0,1]$, with $\xi=1$ corresponding to the 't Hooft–Feynman gauge and $\xi=0$ representing the Landau gauge, where we take the renormalization scale as $\bar{\mu}=T$ with the relation between 4D and 3D parameters as in Refs.~\cite{Croon:2020cgk,Qin:2024idc}, wherein uncertainties induced by the renormalization scale has been shown to negligible at two-loop order~\cite{Croon:2020cgk,Qin:2024idc}. As shown in Fig.~\ref{fig:S3}, the traditional method yields an O(10-27\%) systematic uncertainty in the tunneling exponent $B$, the gauge-invariant bounce action
$\mathcal{B}_{0}+\mathcal{B}_{1}$ is essentially unchanged across the entire parameter range, with numerical variations remaining below 1\%.

\begin{figure}[!htp]
    \centering
\includegraphics[width=0.7\linewidth]{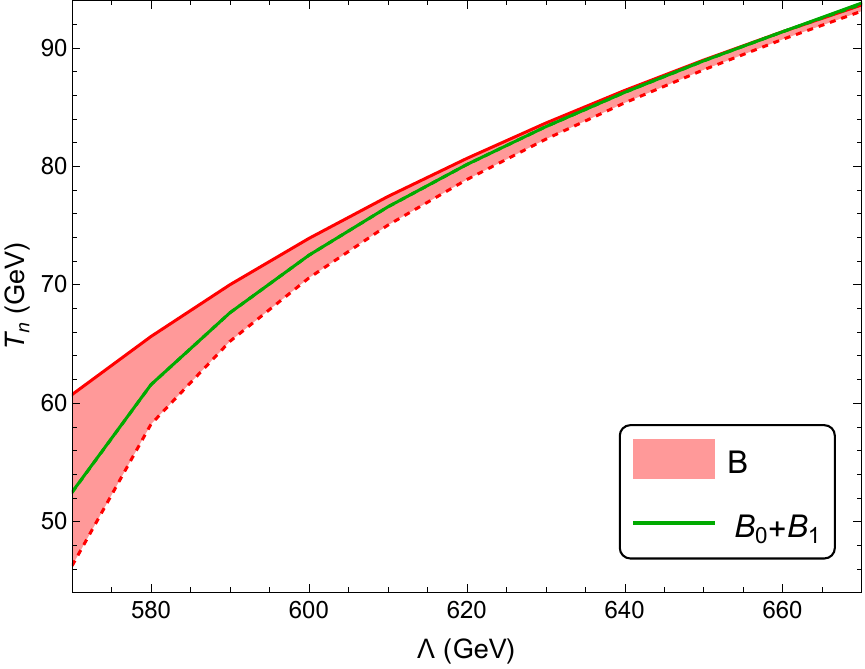}
\quad
\includegraphics[width=0.7\linewidth]{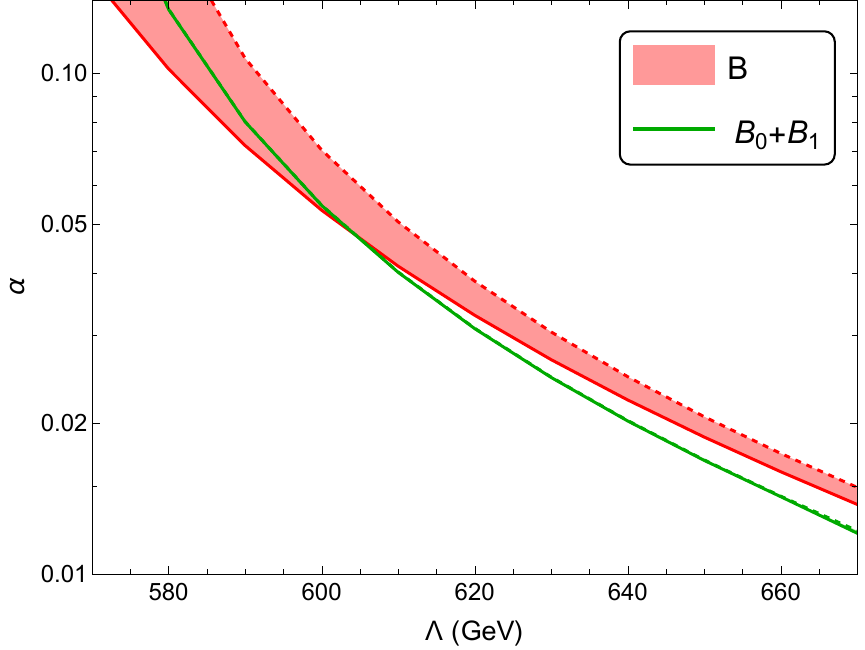}
\quad
\includegraphics[width=0.7\linewidth]{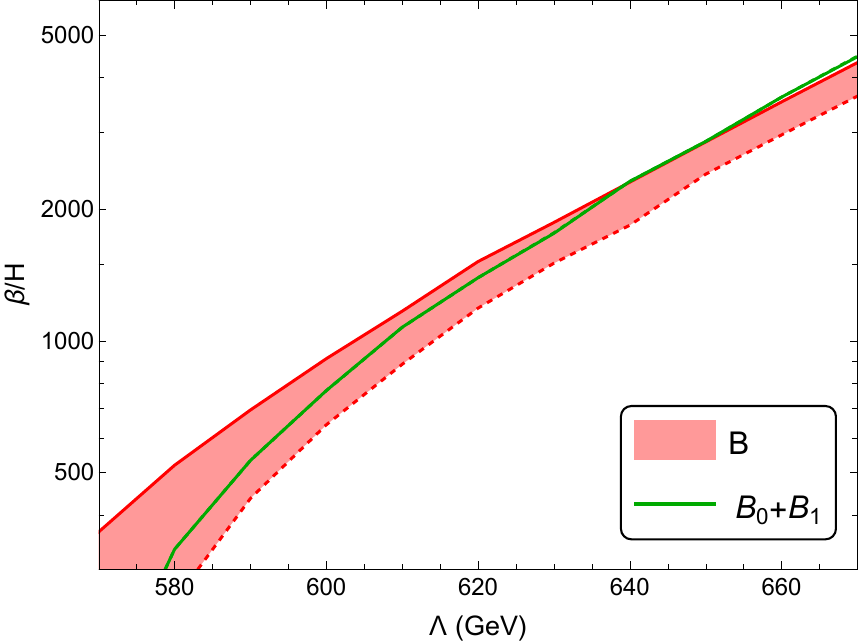}
    \caption{The behaviors of $T_n$(top), $\alpha$ (middle) and $\beta/H$ (bottom) as functions of $\Lambda$. The line-code is the same as Fig.~\ref{fig:S3}. Shaded regions indicate the uncertainty for each observable.}
    \label{fig:Tnalpha&betah}
\end{figure}

\noindent{\it Phase transition parameters and GWs spectra.}
For studying the PT parameters' gauge dependence, we first
calculate the bounce configuration from $\mathcal{S}_0$, and then calculate $\mathcal{B}_1$ with the obtained leading-order bounce configuration as in Refs.~\cite{Hirvonen:2021zej,Lofgren:2021ogg}.
The inverse duration of the PT is obtained from $\mathcal{B}_0+\mathcal{B}_1$, and for the PT trace anomaly we adopt $V_{\rm LO}+V_{\rm NLO}$.
Fig.~\ref{fig:Tnalpha&betah} presents the gauge dependence of PT parameters for the traditional method and the gauge-invariant method at order $\mathcal{B}_0+\mathcal{B}_1$. The temperature $T_n$ and $\beta/H$ increase with $\Lambda$, while $\alpha$ decreases as $\Lambda$ grows. We find that the gauge parameter introduces significant spurious dependence in the traditional approach. In contrast, the key phase transition parameters \(T_n\), \(\alpha\), and \(\beta/H\) are explicitly gauge-independent within our framework: the residual gauge dependence reported in Ref.~\cite{Croon:2020cgk} is completely eliminated, as the bounce action used to calculate these parameters is constructed to be fully gauge-invariant, as demonstrated in the preceding section.

\begin{figure}[!htp]
    \centering
    \includegraphics[width=0.8\linewidth]{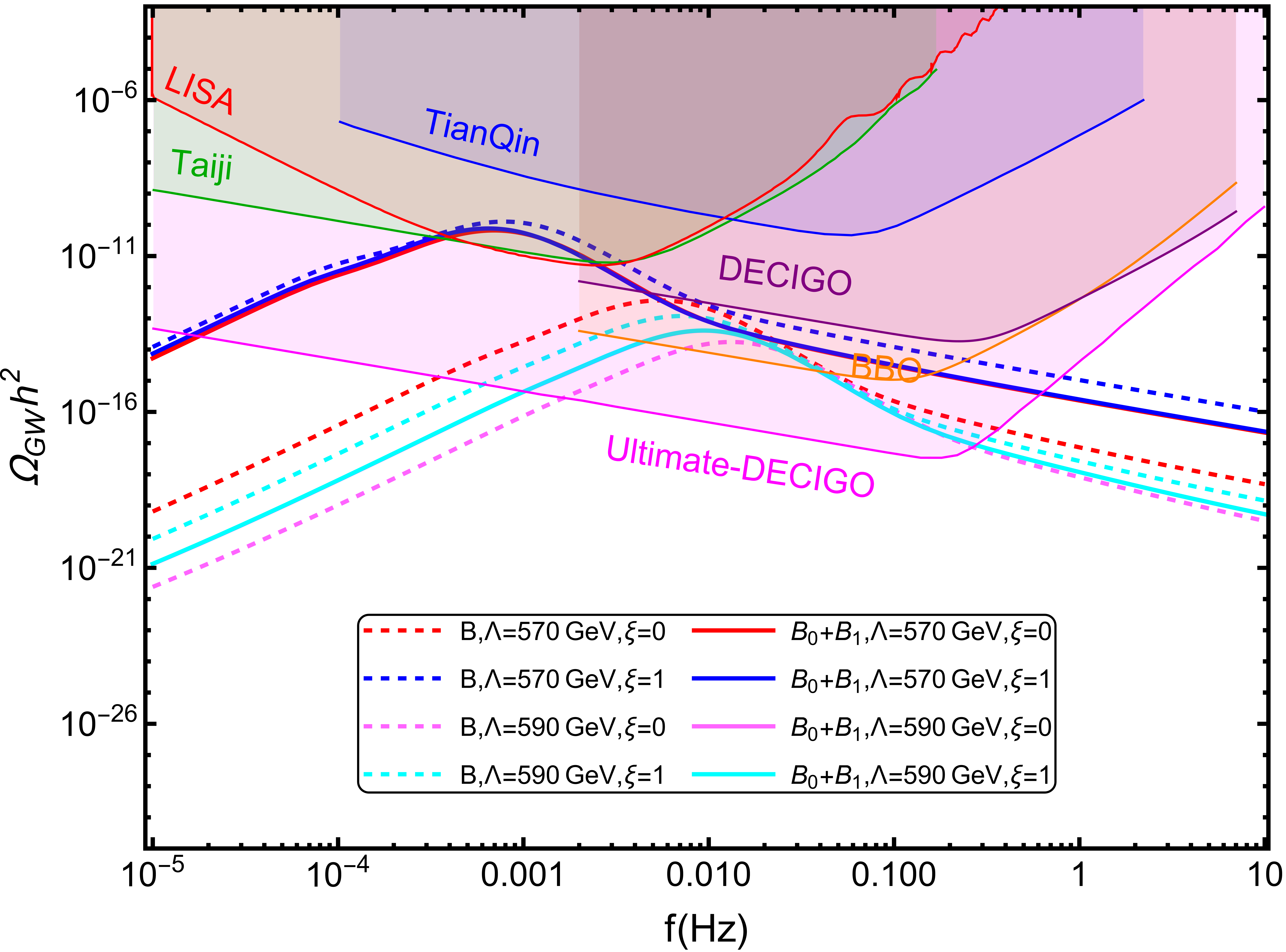}
    \caption{The effect of gauge parameter on GW prediction at $\Lambda=570$ and $590$ GeV. The color region dentes the sensitivity of detectors, including Taiji~\cite{Hu:2017mde,Ruan:2018tsw}, Tianqin~\cite{TianQin:2015yph,Zhou:2023rop}, LISA~\cite{LISA:2017pwj,Baker:2019nia}, BBO~\cite{Crowder:2005nr,Corbin:2005ny,Harry:2006fi} and DECIGO~\cite{Seto:2001qf,Kawamura:2006up,Yagi:2011wg,Isoyama:2018rjb}.}
    \label{fig:gravitational wave}
\end{figure}

Gravitational wave production from first-order cosmological phase transitions receives three primary contributions: bubble collisions~\cite{Huber:2008hg,Caprini:2015zlo,Kamionkowski:1993fg}, sound waves in the plasma~\cite{Hindmarsh:2013xza,Hindmarsh:2015qta,Hindmarsh:2017gnf,Caprini:2019egz,Ellis:2019oqb}, and magnetohydrodynamic (MHD) turbulence~\cite{Caprini:2009yp}. Using our gauge-invariant phase transition parameters, we compute the resulting gravitational wave spectra following the standard phenomenological formulas compiled in Ref.~\cite{Qin:2024idc}. Fig.~\ref{fig:gravitational wave} illustrates the impact of the gauge parameter on these predictions. As expected, the gravitational wave amplitude increases with decreasing new physics scale \(\Lambda\). Crucially, our gauge-invariant \(\mathcal{B}_0+\mathcal{B}_1\) method completely eliminates the spurious gauge dependence in the gravitational wave signal amplitude, resolving a major systematic uncertainty that has affected all prior forecasts. Importantly, for \(\Lambda \lesssim 570\) GeV, the predicted signals fall within the sensitivity bands of the LISA and Taiji space-based interferometers.

\noindent{\it Conclusion and discussion.}
In this Letter, we address the long-standing gauge dependence problem of the bubble nucleation rate for first-order electroweak phase transitions. For the first time, we generalize both the effective potential Nielsen identity and the corresponding wave function renormalization Nielsen identity from the Abelian Higgs model to non-Abelian gauge theories, using the three-dimensional thermal SMEFT as our benchmark framework. Our results are valid up to higher-order terms in \(\xi\) that are consistently neglected in our power-counting scheme, with the Nielsen identity satisfied exactly when retaining all terms up to order \(\sqrt{\xi}\). We decompose the effective potential into leading-order and next-to-leading-order contributions according to the Nielsen identity structure: the LO contribution is manifestly gauge-invariant, while the NLO contribution contains terms of order \(\sqrt{\xi}\) that cancel precisely against the wave function renormalization contribution to the bounce action. Explicit evaluation of the combined Nielsen identities demonstrates that the bounce action calculated using the \(\mathcal{B}_0+\mathcal{B}_1\) method is fully gauge-invariant, and our numerical analysis confirms that this method completely eliminates the spurious gauge dependence of the physical bubble nucleation rate.

We then develop a fully gauge-invariant procedure for extracting key phase transition parameters, and derive the first gauge-invariant prediction of the gravitational wave power spectrum in the SMEFT framework. This eliminates the spurious gauge dependence that has plagued all previous gravitational wave predictions from electroweak phase transitions, removing a major source of systematic uncertainty in signal amplitude forecasts.

Our final results demonstrate that primordial gravitational waves produced in the region with the new physics scale \(\Lambda \lesssim 570\) GeV are within the detection reach of Taiji, LISA, and TianQin. Critically, these detection prospects are free from unphysical gauge artifacts, making them the reliable quantitative forecasts for SMEFT-induced electroweak phase transition signals at future space-based interferometers. We note that, in this regime, additional theoretical uncertainties from the truncation of the high-temperature expansion should also be taken into account in future higher-precision studies~\cite{Qin:2024idc}\footnote{See also Ref.~\cite{Bernardo:2025vkz} for the case of a sextic operator arising in the DR procedure of the Abelian Higgs model within the thermal effective field theory framework.}, which is beyond the scope of this work.

We further note that a gauge-invariant prediction of the critical temperature can be obtained using the first Nielsen identity, as demonstrated in Ref.~\cite{Biekotter:2025npc} for the complex singlet model. However, neglecting the second Nielsen identity associated with the wave function renormalization—which is indispensable for constructing a fully gauge-invariant bounce action—can introduce spurious residual gauge dependence in phase transition parameters, leading to large systematic uncertainties in predictions of gravitational wave spectra from first-order cosmological phase transitions.

Our results establish a three-dimensional EFT model for first-order phase transitions motivated by new physics, which is directly applicable to lattice computations.
This framework therefore enables the reliability of perturbative gauge-independent nucleation rates to be tested against non-perturbative calculations~\cite{Moore:2000jw,Moore:2001vf}, a crucial validation given the current focus on thermal fluctuation effects~\cite{Hirvonen:2024rfg,Hirvonen:2025hqn,Wang:2025ooq,Bian:2025twi}.
The computed nucleation rate provides essential input for simulating a suite of cosmological signatures. Subsequent real-time simulations can explore the generation of primordial magnetic fields~\cite{Di:2020kbw}, the dynamical creation of baryon asymmetry~\cite{Di:2024gsl,Di:2025ncl}, and the detailed spectrum of gravitational waves from fluid motions~\cite{Hindmarsh:2015qta,Hindmarsh:2013xza,Hindmarsh:2017gnf,Cutting:2018tjt,Cutting:2019zws,Cutting:2020nla}. The nucleation rate in the supercooling phase transition of dark sector is also crucial for the GWs detection.~\cite{Kierkla:2025qyz}. Therefore, this synthesis bridges cosmological observations—including magnetic fields and gravitational waves—with collider physics, offering a complementary pathway to explore new physics~\cite{AlvesBatista:2021eeu,Gould:2019qek,LISACosmologyWorkingGroup:2022jok,Bian:2021ini,Caldwell:2022qsj}.

\noindent{\it Acknowledgements. }We are grateful to Michael J. Ramsey-Musolf and Philipp Schicho for helpful discussions on the gauge dependence problem, and Joonas Hirvonen for clarify issues on the calculation of the nucleation rate, and thank Long-Bin Chen, Wen Chen, Feng Feng, Hai Tao Li, Yan-Qing Ma, Wen-Long Sang, and Jian Wang for helpful discussions on the calculation techniques of some two-loop Feynman diagrams.
This work is supported by the National Natural Science Foundation of China (NSFC) under Grants Nos.12322505, 12547101. We also acknowledge Chongqing Talents: Exceptional Young Talents Project No. cstc2024ycjhbgzxm0020 and Chongqing Natural Science Foundation under Grant No. CSTB2024NSCQJQX0022.

\bibliography{references}

\clearpage

\onecolumngrid
\section{Supplemental Material}


In this supplemental material, we collect some of the computational procedures from the main text as well as some additional comparisons, including: the relation between 3D and 4D theory, the $R_\xi$ effective potential, the calculation of factor C and Z are given.


\subsection{From 4D theory to 3D EFT}
The classical Lagrangian density of 4D SMEFT for the study of first-order electroweak phase transition is
\begin{equation}
\label{SMEFT}
\mathcal{L}=\mathcal{L}_{\mathrm{YM}}+\mathcal{L}_{\mathrm{H}}+\mathcal{L}_{\mathrm{F}}+\mathcal{L}_\mathrm{g.f}+\mathcal{L}_{\mathrm{ghost}}\;.
\end{equation}
where one has
\begin{eqnarray}
\mathcal{L}_\mathrm{YM}&& =-\frac{1}{4} W^a_{\mu \nu}W^{a \mu\nu}-\frac{1}{4} B_{\mu\nu}B^{\mu \nu}\,,\\
 \mathcal{L}_\mathrm{H}&&=(D_\mu H)^\dagger (D_\mu H)-V(H)\,,\\
  \mathcal{L}_{\mathrm{F}}&&=\bar{Q}_L i\gamma^\mu D_\mu Q_L+ \bar{t}_R i\gamma^\mu D_\mu t_R\nonumber\\
&&+(-y_t \bar{Q}_L(i \sigma^2)H^* t_R+h.c.)\;,
\end{eqnarray}
with
\begin{equation}
\begin{aligned}
    W^a_{\mu\nu}&=\partial_\mu A^a_\nu-\partial_\nu A^a_\mu+g \epsilon^{abc}A^b_\mu A^c_\nu\,,\\
    B_{\mu\nu}&=\partial_\mu B_\nu-\partial_\nu B_\mu\;,
\end{aligned}
\end{equation}
where $A_{\mu}^{a}$ $(a=1,2,3)$ and $B_{\mu}$ are the SU(2) and U(1) gauge fields, $H$ is the SM Higgs doublet with hypercharge $Y=1$, and $Q_{L}^{T}=(t_{L},b_{L})$ is the left-handed third generation quark doublet. Among the fermions, only the top quark is retained, and the QCD indices are suppressed in the quark sector. The expansion of $H$ reads
\begin{equation}
	H=\frac{1}{\sqrt{2}} \lp \begin{matrix}
		\chi^1(x) + i \chi^2(x) \\
		\phi + h(x) + i \chi^3(x)
	\end{matrix} \rp
\end{equation}
The covariant derivative is defined as
\begin{equation}
	D_{\mu}=\partial_{\mu}-i g \frac{\sigma^{a}}{2}A_{\mu}^{a}-i g^{\prime} \frac{Y}{2}B_{\mu}\;,
\end{equation}
where $\sigma^{a}(a=1,2,3)$ are the Pauli matrices. The Higgs potential is
\begin{equation}
	V(H)=-m^{2}H^{\dagger}H+\lambda\lp H^{\dagger}H\rp^{2}+\frac{1}{\Lambda^2}\lp H^{\dagger}H\rp^{3}\;.
\end{equation}
In perturbation theory, the gauge can be fixed using $R_\xi$-gauge
\begin{equation}
	\begin{aligned}
		\mathcal{L}_\mathrm{g.f}=&-\frac{1}{2 \xi}\lp\partial_\mu A^{a \mu}+i\frac{g}{2} \xi \lp H^\dagger \sigma^a \Phi_0-\Phi_0^\dagger \sigma^a H\rp \rp^2-\frac{1}{2 \xi}\lp \partial_\mu B^\mu+i\frac{g'}{2} \xi\lp H^\dagger \Phi_0-\Phi_0^\dagger H \rp\rp^2\;,
	\end{aligned}
\end{equation}
where
\begin{equation}
	\Phi_0=\frac{1}{\sqrt{2}}\begin{pmatrix}
		0\\
		\tilde{\phi}
	\end{pmatrix}\;,
\end{equation}
with Fadeev-Popov ghost($c^a,\bar{c}^a,c^0,\bar{c}^0$) Lagrangian
\begin{equation}
	\mathcal{L}_{\mathrm{ghost}}=-\begin{pmatrix}
		\bar{c}^a & \bar{c}^0
	\end{pmatrix} \begin{pmatrix}
		M^{ab} & M^a\\
		M^b & M
	\end{pmatrix}\begin{pmatrix}
		c^b\\
		c^0
	\end{pmatrix}\;.
\end{equation}
These matrix elements are:
\begin{equation}
	\begin{aligned}
		M^{ab}&=(\partial^\mu D_\mu^{ab})+g^2\xi[(t^b H)^\dagger(t^a \Phi_0)+(t^a \Phi_0)^\dagger(t^b H)]\;,\\
		M^a&=\frac{g g'}{2}\xi [H^\dagger t^a \Phi_0+(t^a \Phi_0)^\dagger H]\;,\\
		M^b&=\frac{g g'}{2}\xi [(t^b H)^\dagger \Phi_0+\Phi_0^\dagger (t^b H)]=M^a\;,\\
		M&=\partial^2+\frac{g^{'2}}{4}(H^\dagger \Phi_0+\Phi_0^\dagger H)\;,
	\end{aligned}
\end{equation}
where $t^a=\sigma^a/2$ and $D^{ab}_\mu$ is covariant derivative in the adjoint representation,
\begin{equation}
	D^{ab}_\mu=\partial_\mu \delta^{ab}-g f^{abc}A_\mu^c\,.
\end{equation}
Although one needs not relate $\tilde{\phi}$ directly to $\phi$, eventually $\tilde{\phi}$ is identified with $\phi$ to eliminate the mixing between the gauge field and Goldstone mode.

The 3d effective theory, which we can write as,
\begin{equation}\begin{aligned}
   \mathcal{L}^{\mathrm{soft}}_{3d}=&-\frac{1}{4}W^a_{ij}W^a_{ij}-\frac{1}{4}B_{ij}B_{ij}+\frac{1}{2}\lp D_i A^a_0\rp^2+\frac{1}{2}(\partial_i B_0)^2+(D_i H)^\dagger(D_i H)-V^{\mathrm{soft}}_{3d}
\end{aligned}
\end{equation}
where
\begin{equation}
	\begin{aligned}
		W^a_{ij}&=\partial_i A^a_j-\partial_j A^a_i+g f^{abc}A^b_i A^c_j\;,\\
		B_{ij}&=\partial_i B_j-\partial_j B_i
	\end{aligned}
\end{equation}
with
\begin{equation}
\begin{aligned}
D_i A^a_0&=\partial_i A^a_0+g f^{abc}A^b_i A^c_0 \,,\\
   D_i H&=\lp \partial_i-ig \frac{\sigma^a}{2} A^a_i-i g' \frac{Y}{2} B_i\rp H \,.
\end{aligned}
\end{equation}
The scalar potential in the soft scale 3d theory reads
\begin{equation}
\begin{aligned}\label{Vsoft3d}
V^{\mathrm{soft}}_{3d}=&-m_3^2 H^\dagger H+\lambda_3 (H^\dagger H)^2+c_{6,3} (H^\dagger H)^3+\frac{1}{2}m_D^2A^a_0A^a_0+\frac{1}{2}m_D^{'2}B_0^2+\frac{1}{4}\kappa_1(A^a_0A^a_0)^2\\
&+\frac{1}{4}\kappa_2B_0^4+\frac{1}{4}\kappa_3 A^a_0A^a_0B_0^2+h_1 A^a_0A^a_0H^\dagger H+h_2 B_0^2 H^\dagger H+h_3 B_0 H^\dagger A^a_0 \sigma^a H\;,
\end{aligned}
\end{equation}
and together with 3D gauge couplings $g_3$ and $g'_3$. Here $g_3^2$, $g_3^{'2}$ and $\lambda_3$ have dimensions of $[\rm GeV]$ and all the fields have dimensions of $[\rm GeV]^{1/2}$. Upon integrating out the $A_0,B_0$ scalars, we obtain the 3D effective field theory for the SMEFT at the ultra-soft scale, as described in the main text. The Z-factor (and the effective potential) for new physics models can be obtained in {\it DRalgo}~\cite{Ekstedt:2022bff} within the Landau gauge.

\subsection{The calculation of the effective potential}\label{appeff}
At the ultra-soft scale, using the $(R_\xi)$ gauge, which introduces additional topological Feynman diagram contributions compared to the Landau gauge $(\xi=0)$ in Ref.~\cite{Croon:2020cgk}, i.e., the diagrams come from the scalars and ghosts. The corresponding results are listed below:

\begin{equation}
   \mathcal{D}_{SGG}(m_1,m_2,m_3)=-H(m_1,m_2,m_3)\;,
\end{equation}
where
\begin{equation}
    H(m_1,m_2,m_3)=\frac{1}{(4 \pi)^2}\lp \frac{1}{4 \epsilon}+\ln\lp\frac{\bmu}{m_1+m_2+m_3}\rp+\frac{1}{2}\rp\;,
\end{equation}
\be
\begin{aligned}
    (\mathrm{SGG})&=\frac{1}{2}C_{h c_z \bar{c}_Z}^2 \mathcal{D}_{SGG}(m_h,m_{c_Z},m_{c_Z})+\frac{1}{2}C_{h c^+ \bar{c}^-}^2 \mathcal{D}_{SGG}(m_h,m_{c_W},m_{c_W})\,\\
    &+\frac{1}{2}C_{h c^- \bar{c}^+}^2 \mathcal{D}_{SGG}(m_h,m_{c_W},m_{c_W})+\frac{1}{2}C_{G c^+ \bar{c}^-}^2 \mathcal{D}_{SGG}(m_{\chi^0},m_{c_W},m_{c_W})\,\\
    &+\frac{1}{2}C_{G c^- \bar{c}^+}^2 \mathcal{D}_{SGG}(m_{\chi^0},m_{c_W},m_{c_W})-C_{G^+ c^- \bar{c}_Z} \times C_{G^- c^Z \bar{c}^+}\mathcal{D}_{SGG}(m_{\chi^\pm},m_{c_W},m_{c_Z})\,\\
    &-C_{G^+ c_Z \bar{c}^-} \times C_{G^- c^+ \bar{c}_Z}\mathcal{D}_{SGG}(m_{\chi^\pm},m_{c_W},m_{c_Z})\;.
\end{aligned}
\ee
The list of required vertex coefficients reads
\begin{equation}
\begin{aligned}
 &C_{h c_z \bar{c}_Z}=-\frac{g}{2 \cos\theta} m_Z \xi \;,\\
 &C_{h c^+ \bar{c}^-}=C_{h c^- \bar{c}^+}=-\frac{1}{2}g m_W \xi\;,\\
 &C_{G c^+ \bar{c}^-}=C_{G c^- \bar{c}^+}=\frac{i}{2}g m_W \xi \;,\\
 &C_{G^+ c^- \bar{c}_Z}=C_{G^- c^+ \bar{c}_Z}=\frac{1}{2}g m_Z \xi \;,\\
&C_{G^- c^Z \bar{c}^+}=C_{G^+ c_Z \bar{c}^-}=-\frac{g \cos2\theta}{2\cos\theta}m_W \xi \;.
\end{aligned}
\end{equation}

The background field dependent masses are:
\begin{equation}
\begin{aligned}
&h: m_h^2=-m^2+3 \lambda \phi^2+\frac{15}{4}c_6 \phi^4\\
&G: m_{\chi^0}^2=m_G^2+\xi m_Z^2,\,\\
&G^\pm: m_{\chi^{\pm}}^2=m_G^2+\xi m_W^2,\,\\
&W^\pm: m_W^2=\frac{1}{4}g^2\phi^2,\\
&Z: m_Z^2=\frac{1}{4}(g^2+g^{\prime 2})\phi^2,\,\\
&c^\pm: m_{cW}^2=\xi m_W^2,\quad c_Z: m_{cZ}^2=\xi m_Z^2\;.
\end{aligned}
\end{equation}
with $m_G^2=-m^2+\lambda \phi^2+\frac{3}{4}c_6 \phi^4$.
Here, we note that, though other diagrams are the same as in the Landau gauge, the results in the $R_\xi$ gauge are more complicated and lengthy, detailed formulas can be found in the accompanying article.~\cite{Liu:2026ask}.

\begin{figure}[!htp]
    \centering
    \includegraphics[width=0.4\linewidth]{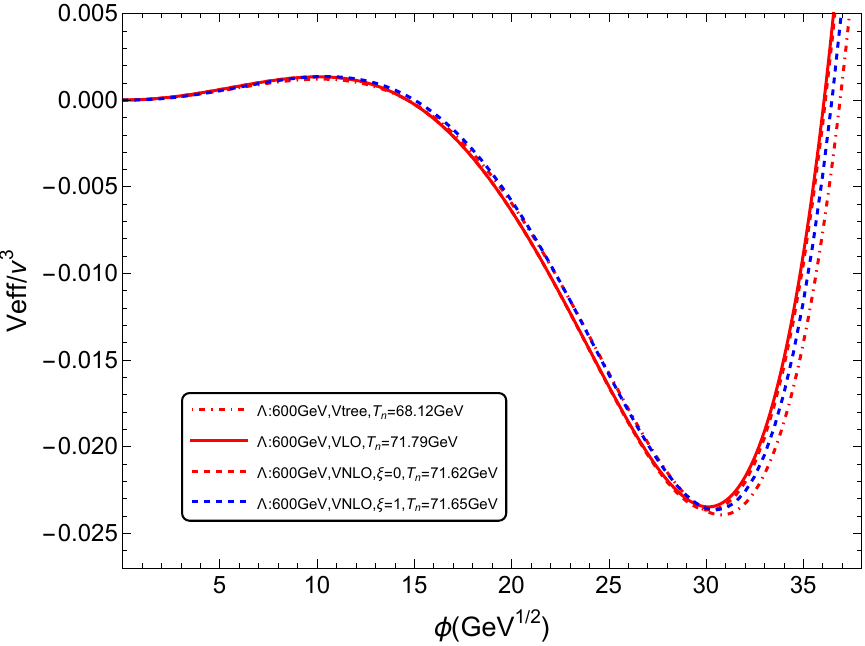}
    \includegraphics[width=0.4\linewidth]{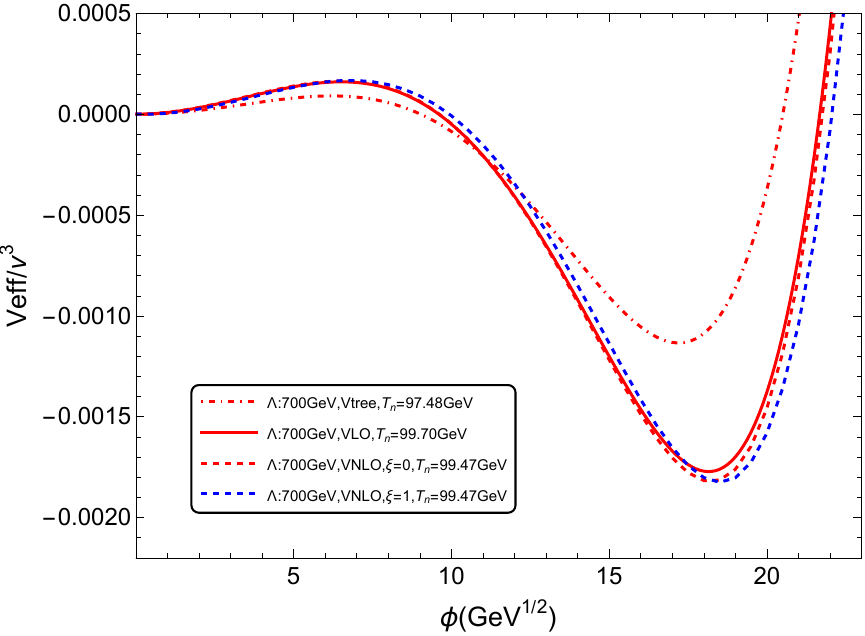}
    \caption{Effective potentials \(V(\phi)\) at tree-level, LO, and NLO, for $\Lambda = 600$GeV (Left) and $\Lambda = 700$GeV (Right). }
    \label{fig:TreeVSLOVSNLO}
\end{figure}

\begin{figure}[!htp]
    \centering
    \includegraphics[width=0.4\linewidth]{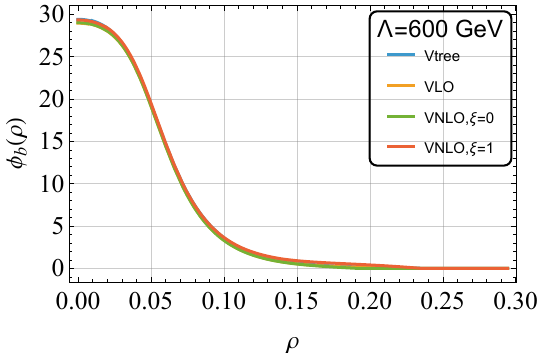}
    \includegraphics[width=0.4\linewidth]{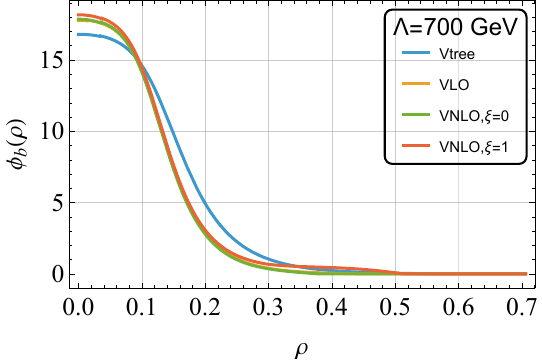}
        \caption{ Bounce solutions at $T_n$ obtained from the tree-level (tree), leading-order (LO), and next-to-leading-order (NLO) effective potentials, for $\Lambda = 600$ GeV (Left) and $\Lambda = 700$ GeV (Right). }
    \label{fig:bouce}
\end{figure}

As illustrated in Figure~\ref{fig:TreeVSLOVSNLO}, the perturbative series for the effective potential exhibits satisfactory convergence at the NLO level. For low new physics scales, exemplified by \(\Lambda=600\) GeV, thermal loop corrections are numerically insignificant. As \(\Lambda\) increases to 700 GeV, however, loop contributions become substantial, and the NLO and leading-order (LO) effective potentials converge toward each other. The dependence on the parameter \(\xi\) introduces only subtle modifications to the overall shape of the effective potential. The bounce solutions computed for these two benchmark points are displayed in Figure~\ref{fig:bouce}, revealing that the discrepancy between LO and NLO bounce profiles becomes pronounced at higher NP scales, whereas the impact of \(\xi\) variations is comparatively modest.

\subsection{C factor of Nielsen Identity}\label{appendix:Cfactor}
The key function for the calculation of the $C$ factor is $K[\phi,\xi]$,
\begin{equation}
	\begin{aligned}
		K_j[\phi(x),\xi]=
        &-\int \diff^4 x i \hbar \bra{0}T \lp\frac{i}{\hbar}\rp^2\left[ \frac{1}{2} \bar{c}^a(x)\lp\partial_\mu W^{a \mu}+g v^a_i \varphi_i\rp igc^b(0) t^b_{jk}\varphi_k(0) \exp \frac{i}{\hbar}S_\mathrm{eff}\right]\ket{0}\,\\
		&-\int \diff^4 x i \hbar \bra{0}T \lp\frac{i}{\hbar}\rp^2\left[ \frac{1}{2} \bar{c}^0(x)\lp\partial_\mu B^\mu+g' v_i \varphi_i\rp ig' c^0(0) n'_{jk}\varphi_k(0) \exp \frac{i}{\hbar}S_\mathrm{eff}\right]\ket{0}\;,
	\end{aligned}
\end{equation}

\begin{figure}[h!]
	\centering
	\includegraphics[width=0.7\linewidth]{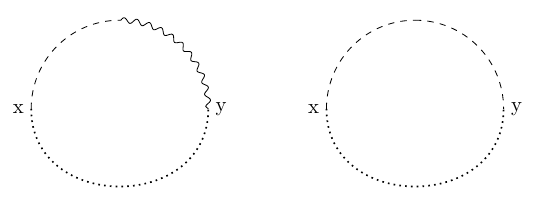}
	\caption{The two graphs that contribute to $C$ at one-loop order.}
	\label{fig:C:one loop}
\end{figure}

The first contribution about $K$ is
\begin{equation}
	\begin{aligned}
		C_i=&-\frac{g}{2}\int_y\braket{c^b(x) t^b_{ij}\varphi_j\bar{c}^a(x)\lp\partial_\mu W^{a \mu}(y)+g v^a_k \varphi_k(y)\rp}-\frac{g'}{2}\int_y\braket{c^0(x) n'_{ij}\varphi_j\bar{c}^0(x)\lp\partial_\mu B^\mu(y)+g' v_k \varphi_k(y)\rp}.
	\end{aligned}
\end{equation}
The corresponding Feynman diagram is shown in Figure \ref{fig:C:one loop}

By using $J(x)=\frac{\delta S}{\delta \phi(x)}$ as an external source\cite{Hirvonen:2021zej}, we have

\begin{equation}
\begin{aligned}
     C&=C_{JG^\pm c^\pm}C_{G\bar{c}^\pm}I_{\mathrm{3d}}(m_{\chi^\pm},m_{c_W})+C_{JG c_Z}C_{G\bar{c}_Z}I_{\mathrm{3d}}(m_{\chi^0},m_{c_Z})\;.
\end{aligned}
\end{equation}

with
\begin{equation}
\begin{aligned}
    I_{3d}(m_1,m_2)&=\int_p \frac{1}{(p^2+m_1^2)(p^2+m_2^2)}=\frac{1}{4 \pi(m_1+m_2)}\;.
\end{aligned}
\end{equation}
Here,
\begin{equation}
	\int_{p} \equiv \mu^{2 \epsilon}\int \frac{\diff^d p}{(2 \pi)^d}\;,
\end{equation}
with $\mu$ being the regularization scale, which has the $\MSbar$ renormalization scale $\bar{\mu}$,
\begin{equation}
	4 \pi \mu^2=e^\gammaE \bar{\mu}^2
\end{equation}
for $d=3-2\epsilon$ in the 3D EFT, and $\gammaE$ is the Euler-Mascheroni constant.

The relevant vertices are:
\begin{eqnarray}
	&C_{J G^\pm c^\pm}=\frac{g}{4}\,,\\
	&C_{G^\pm \bar{c}^\pm}=\frac{1}{2}g \xi \tilde{\phi} \,,\\
	&C_{J G c_Z}\frac{\sqrt{g^2+g^{'2}}}{4}\,,\\
	&C_{G \bar{c}_Z}=\frac{1}{2}\sqrt{g^2+g^{'2}} \xi \tilde{\phi}\,.
\end{eqnarray}
At leading order in our power counting (one can set $m_{\chi^\pm} \to m_{c_W}, m_{\chi^0} \to m_{c_Z}$)
\begin{equation}
    C_{\mathrm{LO}}=\frac{\lp 2 g+\sqrt{g^2+g^{'2}}\rp \sqrt{\xi}}{32 \pi}\,,
\end{equation}
At the soft scale, $A^a_0$ and $B_0$ do not participate in the calculation of the $C$ factor, so the results are similar; only the parameters are changed to the corresponding soft-scale parameters.

\subsection{Z factor}
\label{secZfactor}

\begin{figure}[!htp]
    \centering    \includegraphics[width=0.8\linewidth]{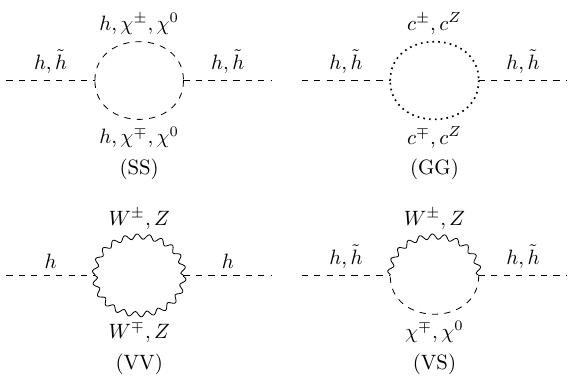}
    \caption{The diagrams contributing to the kinetic tern of effective action.}
    \label{fig:Z factor in 3d with hbar}
\end{figure}

When we calculate the field renormalization factor $Z$, we use
\begin{equation}
    \tilde{\phi} \to \tilde{\phi}+\tilde{h}
\end{equation}
to shift the gauge-fixing background field and treat $\tilde{h}$ as an external auxiliary field that only appears on external legs and does not propagate internally. The field renormalization factor $Z$ comprises the diagrams in Figure~\ref{fig:Z factor in 3d with hbar}.

The field renormalization factor for the scalar field can be computed as
\begin{equation}
	Z=\frac{\partial }{\partial k^2}\lp \Pi_{hh}+\Pi_{h\tilde{h}}+\Pi_{\tilde{h}h}+\Pi_{\tilde{h}\tilde{h}}\rp\;,
\end{equation}
Where $\Pi$ denotes the scalar two-point correlation function listed bellow.

\begin{align}
-\Pi_{hh}=&\frac{1}{2}C_{hhh}^2 I_{SS}(m_h)+\frac{2}{2}C_{hG^+G^-}^2 I_{SS}(m_{\chi^\pm})+\frac{1}{2}C_{hGG}^2I_{SS}(m_{\chi^0})-C_{hZG}^2 I_{VS}(m_Z,m_{\chi^0})\nn\\
    &+\frac{2}{2}C_{hW^+W^-}^2 I_{VV}(m_W)+\frac{1}{2}C_{hZZ}^2 I_{VV}(m_Z)-2 C_{hW^+G^-}\times C_{hW^-G^+} I_{VS}^{HH}(m_W,m_{\chi^\pm})\nn\\
    &-2 C_{hc^+c^-}^2 I_{GG}(m_{c_W})-C_{hc_Zc_Z}^2 I_{GG}(m_{c_Z})\;,\\
-\Pi_{h\tilde{h}}=&\frac{2}{2}C_{hG^+G^-}C_{\tilde{h}G^+G^-}I_{SS}(m_{\chi^\pm})+\frac{1}{2}C_{hGG}C_{\tilde{h}GG}I_{SS}(m_{\chi^0})\nn\\
    &+2 C_{h W^\pm G^\pm} C_{\tilde{h}W^\pm G^\pm}I_{VS}^{H\tilde{H}}(m_W,m_{\chi^\pm})+C_{h Z G}C_{\tilde{h}Z G}I_{VS}^{H\tilde{H}}(m_Z,m_{\chi^0})\nn\\
    &-2 C_{hc^+c^-} C_{\tilde{h}c^+c^-} I_{GG}(m_{c_W})-C_{hc_Zc_Z} C_{\tilde{h}c_Zc_Z}I_{GG}(m_{c_Z})\;,\\
-\Pi_{\tilde{h}\tilde{h}}=&\frac{2}{2}C_{\tilde{h}G^+G^-}^2I_{SS}(m_{\chi^\pm})+\frac{1}{2}C_{\tilde{h}GG}^2I_{SS}(m_{\chi^0})+2 C_{\tilde{h}W^\pm G^\pm}^2I_{VS}^{\tilde{H}\tilde{H}}(m_W,m_{\chi^\pm})\nn\\
&+C_{\tilde{h}Z G}^2I_{VS}^{\tilde{H}\tilde{H}}(m_Z,m_{\chi^0})-2C_{\tilde{h}c^+c^-}^2 I_{GG}(m_{c_W})-C_{\tilde{h}c_Zc_Z}^2I_{GG}(m_{c_Z})\;,
\end{align}
where we have the symmetry $\Pi_{h\tilde{h}}=\Pi_{\tilde{h}h}$. These integration functions of $I_{XY}$ in 3D EFT are given as follows:

\begin{align}
I_{SS/GG}(m)&=\int_{p}\frac{1}{(p^2+m^2)[(p+k)^2+m^2]}=\frac{1}{8 \pi m}+k^2\lp-\frac{1}{96 \pi m^3} \rp+\mathcal{O}(k^4)\;,\\
I_{VV}(m)&=\int_{p}\frac{\left[\delta_{ij}-(1-\xi)\frac{p_i p_j}{p^2+\xi m^2}\right]\left[\delta_{ij}-(1-\xi)\frac{(p+k)_i (p+k)_j}{(p+k)^2+\xi m^2}\right]}{(p^2+m^2)[(p+k)^2+m^2]}\nn\\
&=\frac{\xi^{3/2}+2}{8 \pi m}+k^2 \lp\frac{-9\xi+13 \sqrt{\xi}-10}{96 \pi m^3\lp\sqrt{\xi}+1\rp} \rp+\mathcal{O}(k^4)\;,\\
I_{VS}^{H \tilde{H}}(m_1,m_2)&=\int_p \frac{p_i (p+2 k)_j \left[\delta_{ij}-(1-\xi)\frac{p_i p_j}{p^2+\xi m_1^2}\right]}{(p^2+m_1^2)[(p+k)^2+m_2^2]}\nn\\
		&=-\frac{\xi  \left(m_1^2 \xi +m_2 m_1 \sqrt{\xi }+m_2^2\right)}{4 \pi  \left(m_1 \sqrt{\xi }+m_2\right)}+k^2\lp \frac{ -\xi  \left(3 m_1^2 \xi +6 m_2 m_1 \sqrt{\xi }+2 m_2^2\right)}{12 \pi  \left(m_1 \sqrt{\xi }+m_2\right)^3}\rp+\mathcal{O}(k^4)\;,\\
I_{VS}^{\tilde{H}\tilde{H}}(m_1,m_2)&=\int_p \frac{p_i p_j \left[\delta_{ij}-(1-\xi)\frac{p_i p_j}{p^2+\xi m_1^2}\right]}{(p^2+m_1^2)[(p+k)^2+m_2^2]}\nn\\
    &=-\frac{\xi  \left(m_1^2 \xi +m_2 m_1 \sqrt{\xi }+m_2^2\right)}{4 \pi  \left(m_1 \sqrt{\xi }+m_2\right)}+k^2\lp \frac{ m_1^2 \xi ^2}{12 \pi  \left(m_1 \sqrt{\xi }+m_2\right)^3}\rp+\mathcal{O}(k^4)\;,\\
I_{VS}^{HH}(m_1,m_2)&=\int_{p}\frac{(p+2k)_i (p+2k)_j\left[\delta_{ij}-(1-\xi)\frac{p_i p_j}{p^2+\xi m_1^2}\right]}{(p^2+m_1^2)[(p+k)^2+m_2^2]}\nn\\
		&=-\frac{\xi\lp m_1 m_2 \sqrt{\xi}+m_1^2 \xi+m_2^2\rp}{4 \pi \lp m_1 \sqrt{\xi}+m_2\rp}+k^2 \Bigg( -\frac{4 m_1 m_2 \xi^{3/2}+3m_1^2\xi^2}{12 \pi \lp m_1 \sqrt{\xi}+m_2\rp^3}+\frac{2}{3 \pi (m_1+m_2)}\Bigg)+\mathcal{O}(k^4)\;.
\end{align}

Accordingly, these vertices are:
\begin{align}
   & C_{hhh}=-6 \lambda \phi-15 c_6 \phi^3\,,
\\
& C_{h G G}=C_{h G^+ G^-}=-2 \lambda \phi-3 c_6 \phi^3\,, \\
& C_{h W^+ W^-}=-\frac{1}{2}g^2\phi \,,\\
& C_{h Z Z}=-\frac{1}{2}(g^2+g^{\prime 2})\phi \,,
\\
& C_{h G^+ W^- }=-C_{h G^-W^+ }=\frac{1}{2} g \;,\\
&C_{h W^\pm G^\pm}=-\frac{g}{2} \;,\\
 &C_{Z h G}=-\frac{i}{2}\sqrt{g^2+g^{\prime 2}}\;,\\
&C_{h c_z \bar{c}_Z}=-\frac{g}{2 \cos\theta} m_Z \xi \;,
\\
&C_{h c^+ \bar{c}^-}=C_{h c^- \bar{c}^+}=-\frac{1}{2}g m_W \xi\;,\\
&	C_{\tilde{h} G G}=-\frac{1}{2}\lp g^2+g^{'2} \rp \xi \tilde{\phi}\,,\\
&	C_{\tilde{h} G^+ G^-}=-\frac{1}{2}g^2 \xi \tilde{\phi}\,,\\
&	C_{\tilde{h} W^+ G^-}=-C_{\tilde{h} W^- G^+}=-\frac{i g}{2}\,,\\
& C_{\tilde{h} W^\pm G^\pm}=\frac{g}{2}\,,   \\
&	C_{\tilde{h} Z G}=-\frac{\sqrt{g^2+g^{'2}}}{2}\;,\\
&C_{\tilde{h} c_Z c_Z}=-\frac{1}{4}\lp g^2+g^{'2} \rp \xi \phi\,,\\
&C_{\tilde{h} c^+ c^-}=-\frac{1}{4} g^2 \xi \phi\,.
\end{align}

At leading order in $\lambda\sim g^3$(one can  set $\lambda,m_h \to 0,m_{\chi^\pm} \to m_{c_W}, m_{\chi^0} \to m_{c_Z}$), we have
\begin{equation}
    Z_{\mathrm{NLO}}= -\frac{11 \left(\sqrt{g^2+g^{'2}}+2 g\right)}{48 \pi  \phi }\;.
\end{equation}

\end{document}